\documentclass[aps,prl,reprint,groupedaddress]{revtex4-2}
\usepackage{makecell}
\usepackage{amsfonts}
\usepackage{flafter}
\usepackage{float}
\usepackage{amssymb,graphics,graphicx,subfigure,rotating}

\usepackage{booktabs}
\usepackage{amsmath}
\allowdisplaybreaks[4]

\usepackage{latexsym}
\usepackage{amsmath}
\usepackage[colorlinks=True,linkcolor=blue,anchorcolor=blue,citecolor=blue,CJKbookmarks=true]{hyperref}
\usepackage{xcolor}
\usepackage{natbib}
\usepackage{ctex}

\begin{document}

\title{\textbf{
Entanglement Detection for Two-Qubit and Three-Qubit Pure States via Unitary Transformations and Ancilla State Measurements}}

\author{Yu-Hang Liu$^{1}$}
	\author{Yuan-Hong Tao$^{1}$} \email{taoyuanhong12@126.com}
\author{Yi-Wen Liu$^{1}$}
\author{Ying-Huan Zhou$^{1}$}
	\author{Shao-Ming Fei$^{2}$}
	\affiliation{$^{1}$School of Science, Zhejiang University of Science and Technology, Hangzhou 310023, China\\
		$^{2}$School of Mathematical Science, Capital Normal University, 100048, Beijing, China}
\begin{abstract}
\textbf{\footnotesize{Abstract:}} \footnotesize{Quantum entanglement is the fundamental hallmark of quantum mechanics and a core resource for realizing long-distance quantum communication and scalable linear quantum computing. Accordingly, the precise detection and quantitative quantification of entanglement constitute a foundational and critical problem in quantum information theory. To date, researchers have proposed numerous sufficient conditions for entanglement detection as well as a variety of entanglement measures to characterize the entanglement strength of quantum states; nevertheless, efficient and direct measurement schemes for core entanglement parameters remain underdeveloped. Based on unitary transformations and auxiliary measurements, this paper proposes a set of quantum circuit schemes capable of directly measuring the bipartite concurrence and the tripartite 3-tangle entanglement measure. By introducing auxiliary qubits and constructing specific controlled unitary operations, the proposed scheme maps the analytical expressions of the two entanglement measures onto the measurement probabilities of output states from quantum circuits. It enables efficient and direct quantitative measurement of bipartite and tripartite entanglement without performing full quantum state tomography. This work provides a feasible technical route for the experimental characterization of entanglement properties and lays a groundwork for the practical deployment of multipartite entanglement resources in quantum information processing.}

\textbf{{Keywords:}} {Quantum Circuit, Unitary Transformation, concurrence, 3-tangle}\\
	\end{abstract}
\maketitle
\begin{normalsize}

\section{Introduction}

Quantum entanglement constitutes one of the most fundamental and nonclassical features inherent to quantum mechanics\cite{SC}, serving as the indispensable core resource for cutting-edge quantum technologies including universal quantum computation and secure quantum communication\cite{P,NR}. Nevertheless, the precise detection and quantitative characterization of entanglement remain formidable bottlenecks for both fundamental quantum physics research and practical quantum system applications.
To date, two mainstream strategies have been developed for entanglement detection and quantification, namely quantum state tomography and entanglement witnesses. Quantum state tomography reconstructs the complete density matrix of a target quantum system via a set of mutually complementary projective measurements, enabling the calculation of arbitrary entanglement quantifiers in principle. However, this technique requires massive identical copies of the investigated quantum state for full state reconstruction\cite{JDP,KMD,DPWA}. Critically, the total number of required measurements scales exponentially with the dimension of the quantum system: specifically, approximately $3^n$ measurement settings are demanded for an $n$-qubit quantum state\cite{JDP,TAMM}. Such exponential resource overhead severely deteriorates the scalability of tomography schemes, which greatly restricts their practical deployment in large-scale quantum systems.
For scenarios that only require qualitative verification of entanglement existence rather than full quantitative evaluation, entanglement witnesses provide a resource-efficient alternative. This method identifies entangled states by measuring specific physical observables correlated with entanglement characteristics, which exhibits distinct advantages in reducing measurement resources compared with full state tomography. Even so, entanglement witness schemes still suffer from two inherent drawbacks. First, the construction of optimal witness operators relies heavily on prior knowledge of the target quantum state or state-specific optimization, leading to poor universality for unknown quantum states\cite{SWJ,OCWJ,SPL}. Second, the vast majority of existing witness operators can only accomplish binary judgment of entanglement presence, incapable of quantifying entanglement strength numerically.
Aiming to circumvent the heavy resource cost of full quantum state tomography, tremendous efforts have been devoted to developing direct and low-cost entanglement measurement protocols in recent years. Among these schemes, the controlled SWAP test realizes high-accuracy state comparison and indirect entanglement estimation by introducing auxiliary qubits and coherent controlled unitary operations\cite{RWTW,HRJR,SVT}, opening a novel pathway for efficient entanglement characterization. Compared with tomography and entanglement witnesses, the controlled SWAP test requires far fewer copies of target quantum states, and is competent to quantify the entanglement degree of arbitrary pure states. Subsequent researches further promoted the practical applicability of controlled SWAP tests, extending its application scenarios to multi-qubit systems, experimental optical platforms and weakly mixed quantum states\cite{SOT}.
In addition to experimental extension studies, theoretical innovations based on controlled SWAP tests have also made steady progress. Ref. \cite{JNPM} defined a novel multipartite entanglement quantifier termed collectible entanglement, and verified that this metric can be efficiently estimated via parallel controlled SWAP tests. Combining Bell-basis measurement data, Ref. \cite{JGSN} generalized collectible entanglement from pure states to mixed states via convex roof extension, and further derived a tight lower bound of collectible entanglement for general mixed states based on raw Bell measurement results. On this basis, Ref. \cite{RYSM} demonstrated that controlled SWAP tests are eligible for multipartite mixed-state entanglement detection, and established practical entanglement criteria for multipartite entanglement and genuine multipartite entanglement relying on measured probability distributions.
In this work, we propose a universal direct entanglement detection scheme based on unitary transformations and auxiliary qubit measurements for two-qubit and three-qubit quantum systems. By introducing auxiliary qubits and designing customized controlled unitary operations, we achieve direct and precise calculation of two typical entanglement metrics: the concurrence for two-qubit pure states and the 3-tangle for three-qubit pure states, with corresponding feasible quantum circuits constructed explicitly. The core advantage of our scheme lies in mapping the analytical mathematical expressions of entanglement metrics to the output measurement probabilities of designed quantum circuits, which completely eliminates the necessity of full quantum state tomography.
The detailed workflow of our protocol is illustrated as follows. First, we construct the total initial system state via the tensor product of the target quantum state and its multiple identical copies. Second, three-qubit Toffoli gates are adopted to judge the consistency of qubit states between different subsystems, with auxiliary qubits recording the judgment results. Third, a sequence of single-qubit and two-qubit quantum gates are implemented to encode the analytical formulas of entanglement metrics into the probability amplitudes of the total composite system. Finally, we extract characteristic probability information via projective measurements on auxiliary qubits and partial system qubits, realizing accurate entanglement quantification without full state reconstruction.
We emphasize the experimental feasibility of our scheme from the perspective of existing quantum hardware and software. Three-qubit Toffoli gates have been experimentally realized on multiple mainstream quantum computing platforms, including superconducting circuits, trapped-ion quantum systems and nuclear magnetic resonance quantum processors\cite{ALMM,TKWM,DMWE}. Furthermore, one-step deterministic implementation methods for three-qubit and even arbitrary multi-qubit Toffoli gates have also been reported in recent experimental works\cite{PJRT}. More importantly, Toffoli gates have been integrated as standard built-in modules in prevailing quantum computing programming frameworks such as Qiskit and PennyLane\cite{SJL}. Therefore, our proposed entanglement quantification scheme possesses solid experimental foundations and excellent implementability on current commercial quantum computing devices.
The remainder of this paper is organized as follows. In Sec. II, we briefly review the fundamental theories of concurrence and 3-tangle, and elaborate the structural design of the corresponding quantum circuits. Section III presents the detailed quantum algorithm implementation for two-qubit concurrence quantification. In Sec. IV, we extend the algorithm framework to three-qubit systems and demonstrate the implementation process of 3-tangle measurement. Finally, we conclude the whole work and outlook potential research directions in Sec. V.

\medskip

\vspace{0.2cm}

\section{Preliminaries}

Consider a normalized single-qubit superposition state \(|\psi_1\rangle\)：
\begin{eqnarray*}
|\psi_1\rangle = A_0 |0\rangle + A_1 |1\rangle,
\end{eqnarray*}
where \(A_0, A_1 \in \mathbb{C}\). The probabilities of obtaining measurement outcomes $0$ and $1$ are \(P(|0\rangle) = |A_0|^2\) and \(P(|1\rangle) = |A_1|^2\)\(P(|1\rangle) = |A_1|^2\), respectively.For basis states of multi-qubit systems, we adopt the shorthand notation \(|i\rangle |j\rangle = |ij\rangle\), with \(i,j \in \{0, 1\}\). A general two-qubit state can then be written as \cite{MI}:
\begin{eqnarray*}
|\psi_2\rangle = A_{00} |00\rangle + A_{01} |01\rangle + A_{10} |10\rangle + A_{11} |11\rangle,
\end{eqnarray*}
where \(A_{ij} \in \mathbb{C}\) subject to the normalization condition \(\sum_{i,j} |A_{ij}|^2 = 1\).

For an arbitrary two-qubit state
\begin{eqnarray*}
| \psi \rangle &=& a_{00} |00\rangle_{R}+a_{01} |01\rangle_{R}+a_{10} |10\rangle_{R}+a_{11} |11\rangle_{R} , \\\nonumber
&&{\sum_{ij}|a_{ij}|^2=1}.
\end{eqnarray*}
the composite state $| \psi \rangle$ is entangled if it cannot be expressed as a product state of its subsystems, i.e., $| \psi \rangle \neq | \phi_0 \rangle \otimes | \phi_1 \rangle$ for any pure states $| \phi_0 \rangle$ and $| \phi_1 \rangle$. Its concurrence is defined as \cite{W}:
$$
C_{\psi} = 2 \left| \alpha_{00} \alpha_{11} - \alpha_{01} \alpha_{10} \right|
$$
Concurrence serves as a metric quantifying the degree of entanglement in two-qubit systems, taking values ranging from $0$ to $1$ $(0 < C_2 < 1$). Accordingly, separable (or "product") states yield a concurrence of $0$, while maximally entangled states attain a concurrence of $1$.

For an arbitrary $3$-qubit state
\begin{eqnarray*}
| \psi \rangle &=& a_{000} |000\rangle_{R}+a_{001} |001\rangle_{R}+a_{010} |010\rangle_{R}+a_{011} |011\rangle_{R}\\\nonumber
&&+a_{100} |100\rangle_{R}+a_{101} |101\rangle_{R}+a_{110} |110\rangle_{R}+a_{111} |111\rangle_{R} ,\\\nonumber
&& {\sum_{i}|a_{i}|^2=1}.
\end{eqnarray*}
its 3-tangle is defined as\cite{TF}
\[
\tau_{123} = 4 |d_1 - 2d_2 + 4d_3| 
\]

where
\begin{align*}
d_1 &= a_{000}^2 a_{111}^2 + a_{001}^2 a_{110}^2 + a_{010}^2 a_{101}^2 + a_{100}^2 a_{011}^2, \\[2mm]
d_2 &= a_{000}a_{111}a_{001}a_{110} + a_{000}a_{111}a_{010}a_{101} + a_{000}a_{111}a_{100}a_{011} \\
    &\quad + a_{001}a_{110}a_{010}a_{101} + a_{001}a_{110}a_{100}a_{011} + a_{010}a_{101}a_{100}a_{011}, \\[2mm]
d_3 &= a_{000}a_{011}a_{101}a_{110} + a_{111}a_{001}a_{010}a_{100}.
\end{align*}
The 3-tangle $\tau_{123}$ serves as a crucial metric for quantifying the magnitude of genuine tripartite entanglement in three-qubit systems, with its range of values spanning [0,1]. When $\tau_{123}=0$\(\tau_{123}=0\), no genuine tripartite entanglement exists in the quantum state, which is fully separable. By contrast, $\tau_{123}=1$ corresponds to the maximum amount of genuine tripartite entanglement.

Note that the entanglement structure of the W state can be entirely decomposed into bipartite entanglement; hence it possesses no genuine tripartite entanglement and yields a 3-tangle equal to zero.
\section{Quantum Algorithm for 2-qubit Entanglement Detection}
This section presents the implementation of the quantum algorithm for entanglement detection of $2$-qubit pure states. Based on unitary transformations and auxiliary qubit measurements, this algorithm maps the concurrence that characterizes the entanglement magnitude of $2$-qubit pure states to the measurement probabilities of auxiliary qubits. Entanglement existence judgment and quantitative analysis of entanglement strength can be realized merely via measurement probability results, without reconstructing the full quantum state. Therefore, tomography-free and efficient entanglement detection is achieved for $2$-qubit pure states.
To construct the required quantum circuit, we first prepare one copy of the target state denoted as $|\Psi_2 \rangle$.
The initial composite state of the entire system is then constructed via the tensor product as
\begin{eqnarray*}
	|\Phi_0\rangle &=& |\Psi_1 \rangle \otimes |\Psi_2 \rangle \\\nonumber
\end{eqnarray*}

We introduce a $2$-qubit ancilla system $B$ prepared in the state $|00\rangle_{B}$, along with the operator
\begin{eqnarray*}
W_{j}^{(m)} &=& P^{(m)}_{j} \otimes \sigma^{(x)}_{B^{(j)}} + (I_{j}-P^{(m)}_{j}) \otimes I_{B^{(j)}},
\end{eqnarray*}
where $P^{(m)}_j=|m_{j}\rangle_{R}|m_{j}\rangle_{C}  \;{ _{R}}\langle m_{j}|  {_{C}}\langle m_{j}|$ is the projection operator acting on the $j$-th qubit pair of registers $R$ and $C$, with $m=0,1$, $j=1,2$.
Apply the operator $W_{RCB}^{(1)}=\otimes_{j=1}^{2} W_{j}^{0}W_{j}^{1}$ to $|\Phi_0\rangle |00\rangle_{B}$, obtain
 
\begin{eqnarray*}
	|\Phi_1\rangle &=& W^{(1)}_{RCB}  |\Phi_0\rangle|00\rangle_{B}\\\nonumber
	&=& \big(a_{00} a_{11}|00\rangle_{R} |11\rangle_{C}+a_{01}a_{10} |01\rangle_{R} |10\rangle_{C}\\\nonumber
&&+a_{10}a_{01} |10\rangle_{R} |01\rangle_{C}+a_{11}a_{00} |11\rangle_{R} |00\rangle_{C}\big)|00\rangle_{B}\\\nonumber
&&+|g_1\rangle_{RCB}.
\end{eqnarray*}

Construct the following operator
\begin{eqnarray*}
W_{j} &=& P_{j} \otimes \sigma^{(x)}_{C^{(j)}} + (I_{j}-P_{j}) \otimes I_{C^{(j)}},
\end{eqnarray*}
where $P_j=|1_{j}\rangle_{R}|1_{j}\rangle_{C}  \;{ _{R}}\langle 1_{j}|  {_{C}}\langle 1_{j}|$,  $j=1,2$。
Apply the operator $W_{RC}^{(2)}=\otimes_{j=1}^{2} W_{j}$ to $|\Phi_1\rangle $, obtain
 \begin{eqnarray*}
	|\Phi_2\rangle &=& W^{(2)}_{RC}  |\Phi_0\rangle\\\nonumber
	&=& \big(a_{00} a_{11}|00\rangle_{R} |11\rangle_{C}+a_{01}a_{10} |01\rangle_{R} |11\rangle_{C}\\\nonumber
&&+a_{10}a_{01} |10\rangle_{R} |11\rangle_{C}+a_{11}a_{00} |11\rangle_{R} |11\rangle_{C}\big)|00\rangle_{B}\\\nonumber
&&+|g_2\rangle_{RCB}.
\end{eqnarray*}

Apply the Hadamard operator $W_{R}^{(3)}=H^{\otimes 2}$ to the register $R$
\begin{eqnarray*}
	|\Phi_3\rangle &=& W_{R}^{(3)}  |\Phi_2\rangle\\\nonumber
	&=& \frac{1}{2}\big(2a_{00} a_{11} -2a_{10}a_{01} \big)|11\rangle_{R} |11\rangle_{C}|00\rangle_{B}+|g_3\rangle_{RCB}\\\nonumber
    &=& \frac{2| a_{00} a_{11} -a_{10}a_{01}|}{2}  \frac{\big(a_{00} a_{11} -a_{10}a_{01} \big)}{| a_{00} a_{11} -a_{10}a_{01}|}|11\rangle_{R} |11\rangle_{C}|00\rangle_{B}\\\nonumber
&&+|g_3\rangle_{RCB}\\\nonumber
    &=&\frac{{C_2}}{2} \frac{\big(a_{00} a_{11} -a_{10}a_{01} \big)}{| a_{00} a_{11} -a_{10}a_{01}|}|11\rangle_{R} |11\rangle_{C}|00\rangle_{B}+|g_3\rangle_{RCB}.
\end{eqnarray*}
By measuring the probability of the output state $|11\rangle_{R} |11\rangle_{C}|00\rangle_{B}$ on the registers $R$, $C$ and $B$, as shown in Figure 1, the probability value related to the concurrency degree$P=\frac{{C_2}^2}{4}$。
\begin{figure}[h]
\centerline{\includegraphics[width=0.4\textwidth]{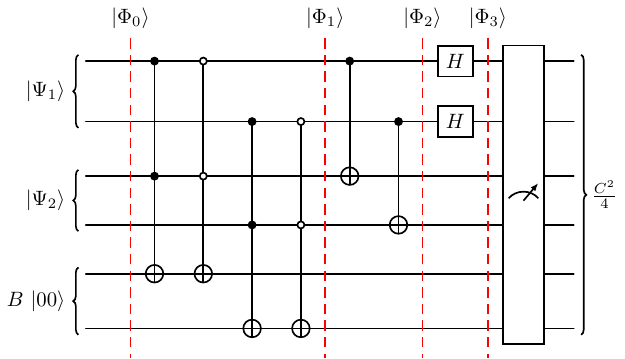}}
\caption{Quantum circuit diagram for 2-qubit entanglement detection}
\end{figure}

\section{Quantum Algorithm for 3-qubit Entanglement Detection}
This section elaborates on the implementation of an entanglement detection algorithm for $3$-qubit pure states. By virtue of unitary operations and ancilla measurements, we map the $3$-tangle (a descriptor of $3$-qubit entanglement) to ancilla measurement probabilities. Genuine tripartite entanglement can be identified and its magnitude quantified merely from measurement data, enabling high-efficiency $3$-qubit entanglement detection without state tomography.

For an arbitrary $3$-qubit pure state $|\Psi \rangle$ given below
\begin{eqnarray*}
| \psi \rangle &=& a_{000} |000\rangle_{R}+a_{001} |001\rangle_{R}+a_{010} |010\rangle_{R}+a_{011} |011\rangle_{R}\\\nonumber
&&+a_{100} |100\rangle_{R}+a_{101} |101\rangle_{R}+a_{110} |110\rangle_{R}+a_{111} |111\rangle_{R} ,\\\nonumber
&&\;\; {\sum_{i}|a_{i}|^2=1}.
\end{eqnarray*}
We divide the calculation of its 3-tangle into the following ten steps.

STEP.1

First, construct the initial state of the overall system
\begin{eqnarray*}
	|\Phi_0\rangle &=& |\Psi_1 \rangle \otimes |\Psi_2 \rangle \otimes |\Psi_3 \rangle \otimes |\Psi_4 \rangle \\\nonumber
\end{eqnarray*}
where each $|\Psi_x \rangle$ denotes a copy of $|\Psi \rangle$, with its information stored in the register$R_x$ for $x=1,2,3,4$。
\\

STEP.2

We introduce two $3$-qubit ancilla systems $B_1$ and $B_2$ initialized to the state $|000\rangle$, together with the operators
\begin{eqnarray*}
W_{mj}^{1} &=& P_{R_1R_2}^{m_j} \otimes \sigma^{(x)}_{B_1} + (I_{R_1R_2}^{m_j}-P_{R_1R_2}^{m_j}) \otimes I_{B_1}, \\\nonumber
W_{mj}^{2} &=& P_{R_3R_4}^{m_j} \otimes \sigma^{(x)}_{B_2} + (I_{R_3R_4}^{m_j}-P_{R_3R_4}^{m_j}) \otimes I_{B_2},
\end{eqnarray*}
where $P_{R_1R_2}^{m_j}=|m_{j}\rangle_{R_1}|m_{j}\rangle_{R_2}  \;{ _{R_1}}\langle m_{j}|  {_{R_2}}\langle m_{j}|, P_{R_3R_4}^{m_j}=|m_{j}\rangle_{R_3}|m_{j}\rangle_{R_4}  \;{ _{R_3}}\langle m_{j}|  {_{R_4}}\langle m_{j}|$, with indices $m=0,1$ and $j=1,2,3$. Here $P_{R_1R_2}^{m_j}$ denotes the projector acting on the $j$-th qubit pair of registers $R_1$ and $R_2$, while $P_{R_3R_4}^{m_j}$ is the projector acting on the $j$-th qubit pair of registers $R_3$ and $R_4$.
Applying operator $W_{R_1R_2R_3R_4B_1B_2}^{(1)}=(\otimes_{j=1}^{3} W_{0j}^{1}W_{1j}^{1}) (\otimes_{j=1}^{3} W_{0j}^{2}W_{1j}^{2})$ to $|\Phi_0\rangle |000\rangle_{B_1}|000\rangle_{B_2}$, obtain  
\begin{eqnarray*}
	|\Phi_1\rangle &=& W^{(1)}_{R_1R_2R_3R_4B_1B_2}  |\Phi_0\rangle|000\rangle_{B_1}|000\rangle_{B_2}\\\nonumber
	&=& \big(|d_1\rangle_{R_1R_2R_3R_4}|000\rangle_{B_1}|000\rangle_{B_2} \\\nonumber
&&+|d_2\rangle_{R_1R_2R_3R_4}|000\rangle_{B_1}|000\rangle_{B_2}\\\nonumber
&&+|d_3\rangle_{R_1R_2R_3R_4}(|100\rangle_{B_1}|100\rangle_{B_2} or |001\rangle_{B_1}|001\rangle_{B_2})\big)\\\nonumber
&&+|g_1\rangle_{R_1R_2R_3R_4B_1B_2}.
\end{eqnarray*}
The ancillas $B_1$ and $B_2$ are utilized here to label the desired terms. The expression $|d_3\rangle_{R_1R_2R_3R_4}$ $(|100\rangle_{B_1}|100\rangle_{B_2} or |001\rangle_{B_1}|001\rangle_{B_2})$ is explained as follows. The state $|d_3\rangle_{R_1R_2R_3R_4}$ comprises multiple types of components: a portion of its terms are labeled by $|100\rangle_ {B_1}|100\rangle_{B_2}$, and the remaining terms are marked by $|001\rangle_ {B_1}|001\rangle_{B_2}$. All subsequent cases follow this pattern analogously.

STEP.3

We now construct the following operator, which restores the states of registers $R_1$ and $R_3$to the basis state $|000\rangle$ for most target terms via ancilla registers $R_2$ and $R_4$.
\begin{eqnarray*}
W_{j}^{1} &=& P_{R_1}^{j} \otimes \sigma^{(x)}_{R_2} + (I_{R_1}^{j}-P_{R_1}^{j}) \otimes I_{R_2}, \\\nonumber
W_{j}^{2} &=& P_{R_3}^{j} \otimes \sigma^{(x)}_{R_4} + (I_{R_3}^{j}-P_{R_3}^{j}) \otimes I_{R_4},
\end{eqnarray*}
where $P_{R_1}^{j}=|0_{j}\rangle_{R_1} \;{ _{R_1}}\langle 0_{j}| $, $P_{R_3}^{j}=|0_{j}\rangle_{R_3}  \;{ _{R_3}}\langle 0_{j}| $, $j=1,2,3$.
Applying operator $W_{R_1R_2R_3R_4}^{(2)}=(\otimes_{j=1}^{3} W_{j}^{1})  ( \otimes_{j=1}^{3} W_{j}^{2})$ to $|\Phi_1\rangle $, obtain
 \begin{eqnarray*}
	|\Phi_2\rangle &=& W^{(2)}_{R_1R_2R_3R_4} |\Phi_1\rangle\\\nonumber
	&=& \big(|d_1\rangle_{R_1R_3} + |d_2\rangle_{R_1R_3}\big)|000\rangle_{R_2}|000\rangle_{R_4}|000\rangle_{B_1}|000\rangle_{B_2}\\\nonumber
&&+|d_3\rangle_{R_1R_3}(|100\rangle_{R_2}|100\rangle_{R_4} or |001\rangle_{R_2}|001\rangle_{R_4})\\\nonumber
&&(|100\rangle_{B_1}|100\rangle_{B_2} or |001\rangle_{B_1}|001\rangle_{B_2})+|g_2\rangle_{R_1R_2R_3R_4B_1B_2}.
\end{eqnarray*} 

STEP.4

We then construct the operator defined below, which fully restores the states of registers $R_2$ and $R_4$ to the basis state$|000\rangle$ for all desired terms by means of the ancillas $B_1$ and $B_2$.
\begin{eqnarray*}
W_{j}^{1} &=& P_{B_1}^{j} \otimes \sigma_{R_2}^{(x)} + (I_{B_1}^{j}-P_{B_1}^{j}) \otimes I_{R_2},\\\nonumber
W_{j}^{2} &=& P_{B_2}^{j} \otimes \sigma_{R_4}^{(x)} + (I_{B_2}^{j}-P_{B_2}^{j}) \otimes I_{R_4}, 
\end{eqnarray*}
where $P_{j}=|1_{j}\rangle_{B_1} \;{ _{B_1}}\langle 1_{j}|,P_{j}=|1_{j}\rangle_{B_2} \;{ _{B_2}}\langle 1_{j}| $ ,$j=1,3$.
Applying operator $W_{R_2R_4B_1B_2}^{(3)}=\otimes_{j=1,3} W_{j_1} \otimes_{j=1,3} W_{j_2}$ to $|\Phi_2\rangle $, obtain
 \begin{eqnarray*}
	|\Phi_3\rangle &=& W^{(3)}_{R_2R_4B_1B_2} |\Phi_2\rangle\\\nonumber
	&=& \big[(|d_1\rangle_{R_1R_3} +|d_2\rangle_{R_1R_3})|000\rangle_{B_1}|000\rangle_{B_2}\\\nonumber
&&+|d_3\rangle_{R_1R_3}(|100\rangle_{B_1}|100\rangle_{B_2} or |001\rangle_{B_1}|001\rangle_{B_2})\big]\\\nonumber
&&|000\rangle_{R_2}|000\rangle_{R_4}+|g_3\rangle_{R_1R_2R_3R_4B_1B_2}.
\end{eqnarray*}

STEP.5

We introduce one $3$-qubit ancilla $B_3$ prepared in the state $|000\rangle$, along with the operator
\begin{eqnarray*}
W_{j}^{m} &=& P_{R_1R_3}^{m_j} \otimes \sigma^{(x)}_{B_1} + (I_{R_1R_3}^{m_j}-P_{R_1R_3}^{m_j}) \otimes I_{B_1}, 
\end{eqnarray*}
where $P_{R_1R_3}^{m_j}=|m_{j}\rangle_{R_1}|m_{j}\rangle_{R_3}  \;{ _{R_1}}\langle m_{j}|  {_{R_3}}\langle m_{j}|, m=0,1$.
Applying operator $W_{R_1R_3B_3}^{(4)}=\otimes_{j=1}^{3} W_{j}^{0}W_{j}^{1} $ to $|\Phi_3\rangle |000\rangle_{B_3}$, obtain
  \begin{eqnarray*}
	|\Phi_4\rangle &=& W^{(4)}_{R_1R_3B_3} |\Phi_3\rangle|000\rangle_{B_3}\\\nonumber
	&=& \big[\big(|d_1\rangle_{R_1R_3}(|000\rangle_{B_3} or |111\rangle_{B_3}) \\\nonumber
&& +|d_2\rangle_{R_1R_3}|X\rangle_{B_3}\big)|000\rangle_{B_1}|000\rangle_{B_2}\\\nonumber
&&+|d_3\rangle_{R_1R_3}|Y\rangle_{B_3}(|100\rangle_{B_1}|100\rangle_{B_2} or |001\rangle_{B_1}|001\rangle_{B_2})\big]\\\nonumber
&&|000\rangle_{R_2}|000\rangle_{R_4}+|g_4\rangle_{R_1R_2R_3R_4B_1B_2B_3}.
\end{eqnarray*}
The ancilla $B_3$ is utilized here to reveal the discrepancy between the corresponding qubit pairs of registers $R_1$ and $R_3$. Moreover, $|X\rangle_{B_3} and |Y\rangle_{B_3}$ do not denote fixed states; instead, they represent a variety of distinct quantum states.

STEP.6

We then construct the operator given below, which equalizes the states of registers $R_1$ and $R_3$ for all desired terms with the help of the ancilla $B_3$.
\begin{eqnarray*}
W_{j} &=& P_{B_3}^{j} \otimes \sigma^{(x)}_{R_3} + (I_{B_3}^{j}-P_{B_3}^{j}) \otimes I_{R_3}, 
\end{eqnarray*}
where $P_{B_3}^{j}=|0_{j}\rangle_{B_3}  \;{ _{B_3}}\langle 0_{j}| $, $j=1,2,3$.
Applying operator $W_{B_3R_3}^{(5)}=\otimes_{j=1}^{3} W_{j}$ to $|\Phi_5\rangle $, obtain
\begin{eqnarray*}
	|\Phi_5\rangle &=& W^{(5)}_{B_3R_3} |\Phi_4\rangle\\\nonumber
	&=& \big[|d_1^{`}\rangle_{R_1R_3}(|000\rangle_{B_3} or |111\rangle_{B_3})|000\rangle_{B_1}|000\rangle_{B_2} \\\nonumber
&&+|d_2^{`}\rangle_{R_1R_3}|X\rangle_{B_3}|000\rangle_{B_1}|000\rangle_{B_2}\\\nonumber
&&+|d_3^{`}\rangle_{R_1R_3}|Y\rangle_{B_3}(|100\rangle_{B_1}|100\rangle_{B_2} or |001\rangle_{B_1}|001\rangle_{B_2})\big]\\\nonumber
&&|000\rangle_{R_2}|000\rangle_{R_4}+|g_5\rangle_{R_1R_2R_3R_4B_1B_2B_3}.
\end{eqnarray*}
For each individual term within $|d_1^{`}\rangle$, $|d_2^{`}\rangle$ and $|d_3^{`}\rangle$, the quantum states of registers $R_1$ and $R_3$are identical inside the same term, whereas the register quantum states corresponding to distinct terms may differ from one another.

STEP.7

We construct the following operator, which fully restores the state of register $R_3$ to the computational basis state $|000\rangle$ for all desired terms via register $R_1$.
\begin{eqnarray*}
W_{j} &=& P_{R_1}^{j} \otimes \sigma^{(x)}_{R_3} + (I_{R_1}^{j}-P_{R_1}^{j}) \otimes I_{R_3}, \\\nonumber
\end{eqnarray*}
where $P_{j}=|1_{j}\rangle_{R_1} \;{ _{R_1}}\langle 1_{j}|$, $j=1,2,3$.
Applying operator $W_{R_1R_3}^{(6)}=\otimes_{j=1}^{3} W_{j_1} $ to $|\Phi_5\rangle $, obtain
\begin{eqnarray*}
	|\Phi_6\rangle &=& W^{(6)}_{R_1R_3} |\Phi_5\rangle\\\nonumber
	&=& \big[|d_1\rangle_{R_1}(|000\rangle_{B_3} or |111\rangle_{B_3}) |000\rangle_{B_1}|000\rangle_{B_2}\\\nonumber
&& +|d_2\rangle_{R_1}|X\rangle_{B_3}|000\rangle_{B_1}|000\rangle_{B_2}\\\nonumber
&&+|d_3\rangle_{R_1}|Y\rangle_{B_3}(|100\rangle_{B_1}|100\rangle_{B_2} or |001\rangle_{B_1}|001\rangle_{B_2})\big]\\\nonumber
&&|000\rangle_{R_2}|000\rangle_{R_3}|000\rangle_{R_4}+|g_6\rangle_{R_1R_2R_3R_4B_1B_2B_3}.
\end{eqnarray*}

STEP.8
We introduce a single $1$-qubit ancilla $B_4$ initialized to the state $|0\rangle$, and construct the following operator such that all terms contained in $|d_1\rangle_{R_1}$ are labeled by $|1\rangle_{B_4}$.
\begin{eqnarray*}
W^{(7)}_{B_3B_4} &=& P_{B_3} \otimes \sigma^{(x)}_{B_4} + (I_{B_3}-P_{B_3}) \otimes I_{B_4}, 
\end{eqnarray*}
where $P_{B_3}=|000\rangle_{R_1} \;{ _{R_1}}\langle 000|+|111\rangle_{R_1} \;{ _{R_1}}\langle 111|$.
Apply the control operator to $|\Phi_6\rangle$
\begin{eqnarray*}
	|\Phi_7\rangle &=& W^{(7)}_{B_3B_4} |\Phi_6\rangle |0\rangle_{B_4}\\\nonumber
	&=& \big[\big(|d_1\rangle_{R_1}(|000\rangle_{B_3} or |111\rangle_{B_3})|1\rangle_{B_4} \\\nonumber
&& +|d_2\rangle_{R_1}|X\rangle_{B_3}|0\rangle_{B_4}\big)|000\rangle_{B_1}|000\rangle_{B_2}\\\nonumber
&&+|d_3\rangle_{R_1}|Y\rangle_{B_3}(|100\rangle_{B_1}|100\rangle_{B_2} or |001\rangle_{B_1}|001\rangle_{B_2})|0\rangle_{B_4}\big]\\\nonumber
&&|000\rangle_{R_2}|000\rangle_{R_3}|000\rangle_{R_4}+|g_7\rangle_{R_1R_2R_3R_4B_1B_2B_3B_4}.
\end{eqnarray*}

STEP.9

Act the Hadamard operator $W_{R_1B_1B_2B_3B_4}^{(8)}= H^{\otimes 3}_{R_1} H^{\otimes 2}_{B_1} H^{\otimes 2}_{B_2}H^{\otimes 3}_{B_3}H_{B_4} $ on $|\Phi_7\rangle$, and obtain
\begin{eqnarray*}
	|\Phi_8\rangle &=& W_{R_1B_1B_2B_3B_4}^{(8)} |\Phi_7\rangle \\\nonumber
	&=& \frac{4\big(2d_2-d_1-4d_3 \big)}{2^{\frac{11}{2}}}
|000\rangle_{R_1}|000\rangle_{R_2}|000\rangle_{R_3}|000\rangle_{R_4}\\\nonumber
&&|101\rangle_{B_1}|000\rangle_{B_2}|000\rangle_{B_3}|1\rangle_{B_4}
+|g_7\rangle_{R_1R_2R_3R_4B_1B_2B_3}.
\end{eqnarray*}
\medskip

STEP.10

We perform a measurement on the composite system $|\Phi_8\rangle$. If the measurement outcome reads
$|000\rangle_ {R_1}|000\rangle_{R_2}|000\rangle_{R_3}|000\rangle_{R_4}|101\rangle_{B_1}
|000\rangle_{B_2}|000\rangle_{B_3}|1\rangle_{B_4}$
the measurement probability associated with the 3-tangle can be directly extracted as $P=\frac{{\tau}^2}{2^{11}}$, as illustrated in Figure 2.
\vspace{0.2cm}

\begin{figure}[h]
\centerline{\includegraphics[width=0.5\textwidth]{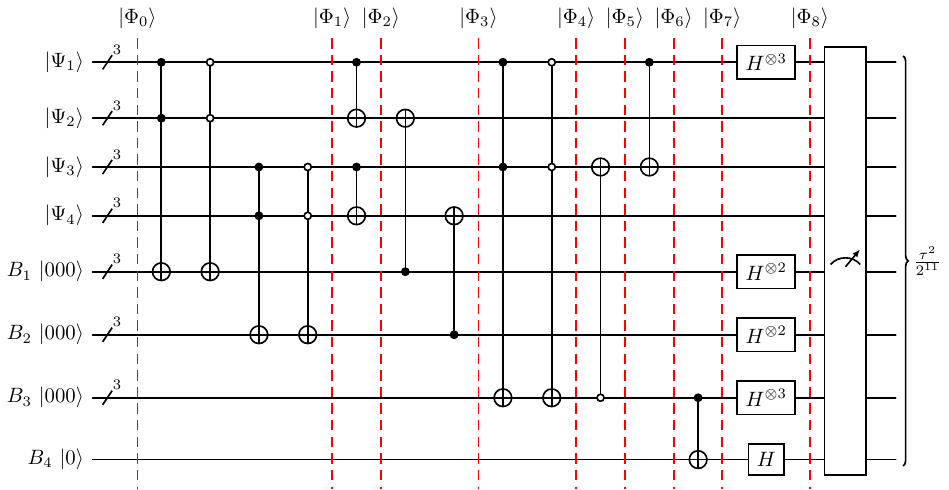}}
\caption{Quantum circuit diagram for 3-qubit entanglement detection}
\end{figure}

\section{Conclusion}
We present two separate quantum algorithms built upon unitary transformations and auxiliary measurements. The first algorithm directly evaluates the concurrence for two-qubit pure states, while the second computes the 3-tangle for three-qubit pure states, with full quantum state tomography entirely avoided. Our protocol prepares multiple identical copies of the target state, incorporates auxiliary qubits and multi-qubit Toffoli gates, and maps the analytical formulas of the corresponding entanglement measures to measurement probabilities of designated output states, which allows efficient, direct readout of entanglement quantifiers.
This work delivers a feasible new quantum computational framework for entanglement quantification. Future research will extend the proposed scheme to mixed states and larger multi-qubit systems, and explore its practical implementation in various quantum information protocols.


\vspace{0.2cm}

\textbf{Acknowledgments}

This work is being supported by the following: the National Natural Science Foundation of China (NSFC) under Grants 11761073, 12075159, and 12171044; the Academician Innovation Platform of Hainan Province.

\vspace{0.2cm}

\end{normalsize}

\end{document}